\documentclass[11pt,a4paper]{article}
\pdfoutput=1
\usepackage{jheppub}

\usepackage{amsmath}
\usepackage{amssymb}
\usepackage{graphicx}
\usepackage{hhline}
\usepackage{textcomp}
\makeatletter
\newcommand{\bea}{\begin{eqnarray}}
\newcommand{\eea}{\end{eqnarray}}

\newcommand{\be}{\begin{equation}}
\newcommand{\ee}{\end{equation}}
\usepackage[active]{srcltx}

      \vspace*{0.2cm}

\title{Solvable self-dual impurity models}

\author[a]{C. Adam}
\author[b]{K. Oles}
\author[c,d]{J. M. Queiruga}
\author[b]{T. Romanczukiewicz}
\author[b]{A. Wereszczynski}

\affiliation[a]{Departamento de F\'isica de Part\'iculas, Universidad de Santiago de Compostela and Instituto Galego de F\'isica de Altas Enerxias (IGFAE) E-15782 Santiago de Compostela, Spain}
\affiliation[b]{Institute of Physics, Jagiellonian University, Lojasiewicza 11, Krak\'{o}w, Poland}
\affiliation[c]{Institute for Theoretical Physics, Karlsruhe Institute
of Technology (KIT), 76131 Karlsruhe, Germany}
\affiliation[d]{Institute for Nuclear Physics, Karlsruhe Institute of Technology (KIT), Hermann-von-Helmholtz-Platz 1,
D-76344 Eggenstein-Leopoldshafen, Germany}

\abstract{
We find a family of (half) self-dual impurity models such that the self-dual (BPS) sector is exactly solvable, for any spatial distribution of the impurity, both in the topologically trivial case and for kink (or antikink) configurations. This allows us to derive the metric on the corresponding one-dimensional moduli space in an analytical form. Also the generalized translational symmetry is found in an exact form. This symmetry provides a motion on moduli space which transforms one BPS solution into another. Finally, we analyse exactly how vibrational properties (spectral modes) of the BPS solutions depend on the actual position on moduli space.

These results are obtained both for the nontrivial topological sector (kinks or antikinks) as well as for the topologically trivial sector, where the motion on moduli space represents a kink-antikink annihilation process.

}

\begin{document}

\maketitle
\flushbottom
    
      
\section{Introduction}
The recently discovered Bogomolny-Prasad-Sommerfeld (BPS)-impurity models (or self-dual impurity models - we will use both names interchangeably)  \cite{BPS-imp}, \cite{BPS-imp-1} possess features which make them a very attractive
theoretical tool for the study of dynamical aspects of topological solitons. In contrast to the usual (1+1) dimensional solitonic models (with or without impurity), they
 possess a nontrivial moduli space, i.e., a space of energetically equivalent static solutions which are solutions of a pertinent Bogomolny equation 
 and, therefore, saturate the corresponding topological bound. Of course, as all configurations have exactly the same energy, there is no static force 
 between the static BPS soliton and the impurity.  Importantly, however, the form of the solution changes at different positions on the moduli space, i.e., for
 BPS solitons with different distances from the impurity. There is, in fact, a transformation called a generalized translation which transforms one BPS solution into 
 another. This transformation is a symmetry of the Bogomolny equation but not of the full action (the full Euler-Lagrange (EL) equation). As a consequence, the low energy dynamics 
 of the BPS-impurity solution follows a (nontrivial) geodesic flow on moduli space. As a consequence of this nontrivial symmetry, the spectral structure of 
 the BPS solution depends on its position on moduli space. In other words, a BPS solution vibrates differently, depending on its distance from the impurity. 
 This allows us to analyze interactions between the BPS soliton and the excited modes in a very clear set-up. Note that, in the usual solitonic models in 
 (1+1) dimensions, the moduli space is trivial. It consists of translationally equivalent BPS solutions (one-soliton solutions) with a fixed spectral structure which does not depend on its position on moduli space.
 On the other hand, the standard ways to couple impurities (see for example \cite{impurity-1}-\cite{impurity-3}) destroy the self-dual (BPS) sector completely. Hence, solitons are attracted or repelled by the impurity, implying a preferred distance between the soliton and the impurity. As a result, no moduli space survives. 

As a consequence,
the BPS-impurity models in one spatial dimension reveal striking similarities to higher dimensional BPS (self-dual) 
theories, like the Abelian Higgs model at critical coupling, or the self-dual t'Hooft-Polyakov monopoles. For example, BPS vortex solutions at different positions on the corresponding moduli space have different shapes and different 
spectral structure (the spectral structure has been calculated explicitlty only for cylindrical vortices \cite{vortex-1}-\cite{vortex-3}). Therefore, the self-dual impurity models in (1+1) dimensions (rather than the usual
 solitonic models without impurities) may serve as a laboratory for higher dimensional self-dual 
 (BPS) theories, especially as far as dynamical properties are concerned. Indeed, they reproduce the 
 usual geodesic flow on moduli space which describes the low speed dynamics of BPS solitons. Moreover, 
 they allow for an analytical study of effects beyond the geodesic approximation like, e.g., an 
 interaction of BPS solitons with internal (oscillating) modes, which can be crucial for a better 
 understanding of the 
 full dynamics as well as quantum properties (see for example the role of the vibrational modes in the 
 semiclassical quantization of the Skyrme model \cite{Sk-vib-1}-\cite{Sk-vib-2}).

It this work, we present a type of BPS-impurity models which, while keeping all the main properties of the BPS-impurity theories unchanged, are {\it analytically solvable} in the BPS regime for any impurity. Furthermore, the metric on the moduli space and, therefore, the geodesic flow, as well as the effective potential in the spectral problem can be determined in an exact manner. This will allow us to understand in an analytical way how the motion on the moduli space (generalized translation) is reflected on the level of the spectral properties of BPS solitons. Thus, we provide the simplest (solvable and (1+1) dimensional) example of a model with a nontrivial moduli space. 

\vspace*{0.1cm} 

It should be underlined that, although the assumed form of the model requires a quite specific coupling between the impurity (which, on the other hand, can have any spatial distribution) and the pre-potential, the same qualitative features are shared by all other self-dual impurity models. The coupling considered here is chosen to render analytical computations possible. In addition, this coupling enlarges the acceptable class of impurities, which now do not have to be $L^2$ integrable. Instead, our impurity can approach a constant value or even diverge at spatial infinity. This opens a new avenue of applications of the self-dual impurity models for a better understanding of the dynamics of annihilation processes in kink-antikink scattering. 

Indeed, we show that such a non-localized impurity can lead to a theory whose BPS sector consists of topologically trivial solutions. The corresponding {\it moduli space describes the annihilation} (creation) of an asymptotically infinitely separated soliton-antisoliton pair. As they are solutions of a Bogomolny equation, there is no static force between the kink and the antikink. Again, the moduli space metric as well as the effective potential of the spectral problem can be found analytically. This may provide a chance to  analytically understand the role of internal modes in the annihilation of solitons. 


\section{Static BPS solutions}
It has been shown recently that for a given (1+1) dimensional scalar field theory with topological solitons (and therefore with a potential with at 
least two vacua), there are infinitely many couplings to an impurity $\sigma=\sigma(x)$ such that half of the BPS sector of the original model keeps the 
BPS property \cite{BPS-imp-2}. That is to say, after the coupling with the impurity, either the kink or the antikink (but never both) is still a solution of a first order differential equation, the so-called Bogomolny equation. In the 
limit $\sigma \rightarrow 0$ all these extensions reproduce the original scalar theory. 

In the present work we want to consider the following BPS-impurity model
\be
 E=\int_{-\infty}^\infty  dx \left[\frac{1}{2}\phi_t^2+ \frac{1}{2} \phi_x^2  + W^2 + \sigma^2 W^2 +\sqrt{2} \sigma W \phi_x + 2 W^2 \sigma \right] ,
 \ee
 which in the limit $\sigma =0$ gives the usual scalar model with a potential $U(\phi)=W^2(\phi)$ (with two vacua $\phi^+>\phi^-$). Note that this BPS-impurity 
 theory differs in some details from the previously analyzed examples \cite{BPS-imp}, \cite{BPS-imp-1}, but the crucial properties, discussed briefly above, remain 
 unchanged. 
 
In order to find a Bogomolny equation, we rewrite the static energy as
 \bea
 E=\int_{-\infty}^\infty  dx \left[ \frac{1}{2} \left( \phi_x +\sqrt{2} W \right. \right. &+& \left. \sqrt{2} \sigma W \right)^2  - \left. \sqrt{2} W \phi_x \right] \nonumber \\
 &\geq&  - \sqrt{2} \int_{-\infty}^\infty dx W \phi_x = \sqrt{2} Q \int_{\phi_-}^{\phi_+}W d\phi
 \eea
where the bound is saturated if the following Bogomolny equation holds
\be
 \phi_x +\sqrt{2} W + \sqrt{2} \sigma W=0. \label{BOG-x}
\ee
Here $Q$ is the topological charge $Q=(\phi(x=\infty) - \phi(x=-\infty))/(\phi^+-\phi^-)$. One may easily verify that the Bogomolny equation 
implies the full equation of motion. Note also that, as always happens for the BPS-impurity models, we have only one Bogomolny equation 
which means that only one soliton (either the kink or the antikink) belongs to the BPS sector, while its charge conjugate partner only solves the full EL equations 
and does not saturate the topological bound. Furthermore, there is no restriction on the spatial distribution of the impurity. 

A peculiarity of this particular class of BPS-impurity models consists in the fact that the Bogomolny equation can be brought into the no-impurity form. Namely, consider a coordinate transformation $y=y(x)$ such that
\be
\frac{dy}{dx} = 1+\sigma(x) \;\; \Rightarrow \;\; y(x)=x+\int_{-\infty}^x \sigma(x') dx' \equiv x+\Delta_\sigma(x)
\ee
where $\Delta_\sigma(x)$ is a coordinate shift due to the impurity.  Then, the Bogomolny equation is just the Bogomolny equation of the no-impurity limit (in the new coordinate $y$)
\be
\phi_y +\sqrt{2} W=0 \label{BOG-y}
\ee
which enjoys the trivial translational symmetry $y \rightarrow y+y_0$, $y_0\in \mathbb{R}$. Now, we can derive the form of the generalized translation analogous to \cite{BPS-imp-1}, which
acts on the BPS solution in terms of the $x$ variable. In the present case, it is a pure coordinate transformation $x\rightarrow \tilde{x} (x,x_0)$, where $x$ and $\tilde{x}$ are related by
\be
\tilde{x}+ \Delta_\sigma(\tilde x) = x+x_0 + \Delta_\sigma(x). \label{transl}
\ee
As we see, the coordinate transformation depends on the particular form of the impurity, i.e., on $\Delta_\sigma (x)$. The 
transformation (\ref{transl}) is an exact, global version of the previously identified infinitesimal 
generalized translations (which, however, in \cite{BPS-imp-1} are not pure coordinate transformations). Obviously, the Bogomolny equation in the new variable (\ref{BOG-y}) is 
solvable by an integration, which ensures the solvability of this self-dual impurity theory.

The BPS sector still possesses the two constant topologically trivial solutions (lumps) $\phi=\phi^\pm$, which 
correspond to the vacua of the non-impurity 
model. As they saturate the topological bound, their energy must be 0. Obviously, the generalized 
translation maps the lump solution into itself, as 
expected on general grounds, see \cite{BPS-imp-1}. As we will see below, despite their simplicity these lump states are not trivial and do 
contribute 
to the dynamics and spectral properties of the solutions. 

Let us observe that the static energy written in the variable $y$ is {\it not} completely equivalent to 
the no-impurity case. Indeed,
\be
E=\int_{-\infty}^\infty  dy \left(1+\tilde{\sigma} (y) \right) \left( \frac{1}{\sqrt{2}} \phi_y +W \right)^2 + {\mbox{boundary term}} \label{E-y}
\ee
where $\tilde{\sigma}(y) = \sigma (x(y))$ is the impurity expressed in the $y$ coordinate. Note that the energy density is proportional to the Bogomolny equation squared. It is therefore a trivial observation that any solution of the
Bogomolny equation $\phi_y +\sqrt{2}W=0$ obeys the full equation of motion. The energy functional (\ref{E-y}) cannot be written as 
$E=\int_{-\infty}^\infty  dy \left(1+\tilde{\sigma} (y) \right) \left( \frac{1}{2} \phi_y^2 +W^2 \right)$ i.e., with the cross term omitted, because this term is multiplied by the impurity function and is not a pure boundary term. Observe also that in the usual, $\tilde{\sigma}=0$ case, the full static equation of motion can be once integrated to a constant pressure equation $\frac{1}{2}\phi_y^2 +W^2=P$ with $P \in \mathbb{R}_+$ being a pressure. In other words, any static solution is a solution of this constant pressure generalization of the Bogomolny equation. This allows, for example, for the construction of solitons in a box (or on a circle $\mathbb{S}^1$ for a periodic pre-potential). Here, on the contrary, solutions of the constant pressure equation do not solve the equation of motion unless $P=0$. Unfortunately, we do not know any generalisation of the Bogomolny equation (\ref{BOG-y}) (or (\ref{BOG-x})) which would be equivalent to the static equation of motion. A related question is whether it is possible to define the self-dual impurity model on a circle (which for topologically nontrivial solutions, of course, requires a periodic impurity). Note that for the original non-impurity model on $\mathbb{S}^1$, solitonic solutions do not saturate the pertinent topological bound as they obey the Bogomolny type equation with a non-zero pressure (here related to the radius of $\mathbb{S}^1$). Hence, strictly speaking, they are not of the BPS type. On the other hand, if put on a circle, they obviously enjoy the usual translation symmetry (with the periodic boundary condition), forming a trivial moduli space.
\begin{figure}
\hspace*{2.0cm}
\includegraphics[height=6.5cm]{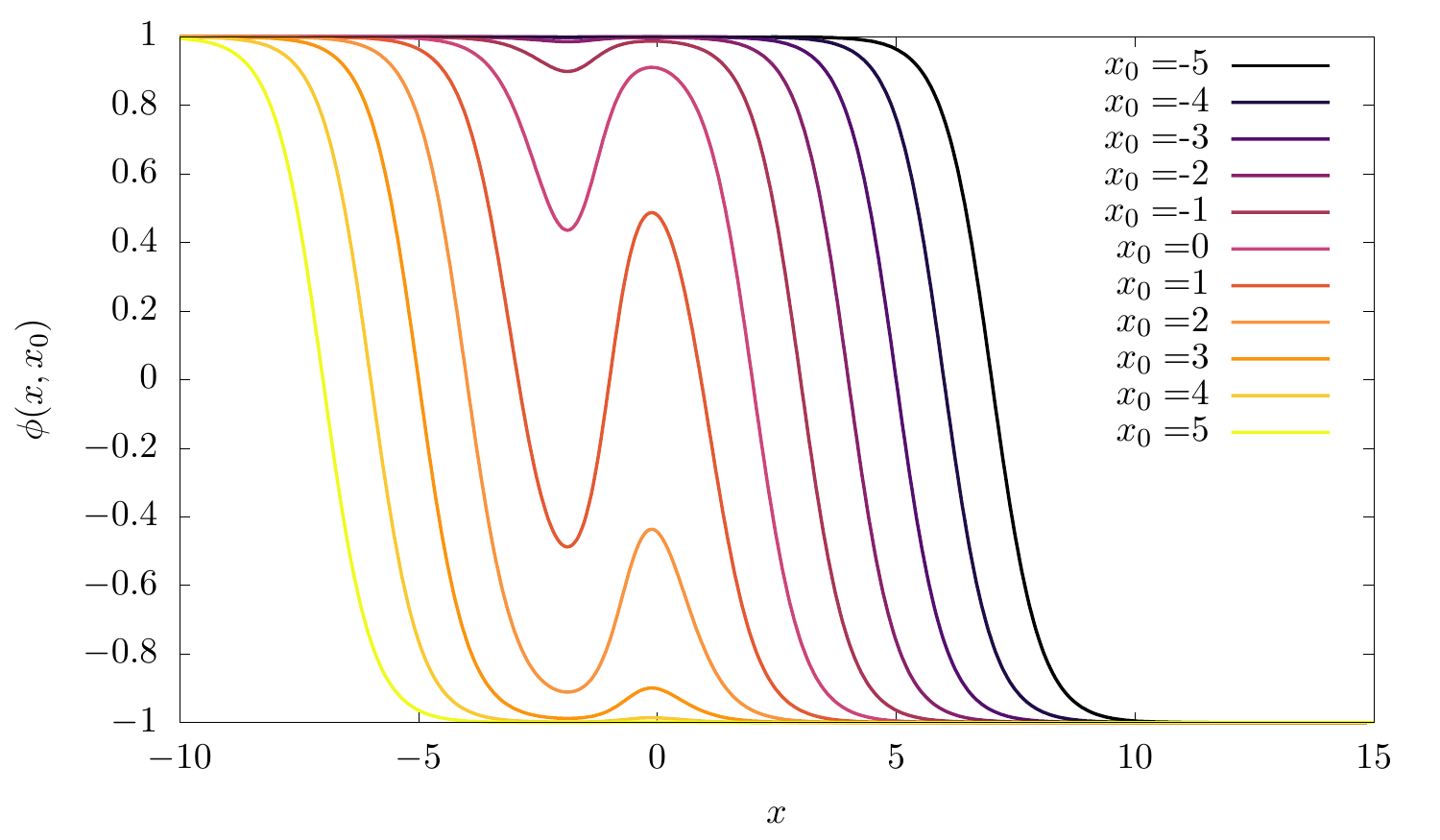}
\caption{Profile of the BPS antikink for the $\phi^4$ BPS-impurity model (\ref{phi4-sol}) with $\alpha=-2$ for different values of the moduli space parameter $x_0$.}
\label{profiles}
\end{figure}

\vspace*{0.2cm}

As an example, let us consider $\phi^4$ theory, that is, we take $W=(1-\phi^2)/\sqrt{2}$. Furthermore, 
we assume the previously considered, exponentially localized form of the 
impurity $\sigma(x)=\alpha/ \cosh^2 x$, where $\alpha \in \mathbb{R}$ measures the strength of the impurity \cite{BPS-imp-1} (see also \cite{impurity-3}). Then, the Bogomolny equation (\ref{BOG-y}) formally has an antikink solution in the $y$ coordinate, which is the antikink of the pure $\phi^4$ theory
\be
\phi(y)_{BPS}=-\tanh (y +y_0).
\ee
Hence, the BPS antikink in the self-dual impurity model is 
\be
\phi(x)_{BPS}=- \tanh (x + \Delta_\alpha (x) +x_0) \label{phi4-sol}
\ee
where the coordinate shift reads
\be
\Delta_\alpha (x)=\alpha \left( \tanh x +1 \right) .
\ee
Its value changes from $\Delta_\alpha(x=-\infty)=0$ to $\Delta_\alpha (x=\infty)=2\alpha$. The moduli space may be parametrized by $x_0\in \mathbb{R}$. In Fig. \ref{profiles} we present profiles of the BPS antikink with $\alpha=-2$ for several values of the moduli space parameter $x_0$ which represents different positions of the soliton relatively to the impurity. The impact of the impurity on the soliton is visible only when the soliton is in the vicinity of the impurity. Otherwise, the profile tends to the usual $\phi^4$ antikink solution. 

Note that for such a strong negative impurity (i.e., for $\alpha =-2$),  the position of the zero of the field (i.e., the position $a$ such that $\phi_{BPS}(a)=0$) cannot be used globally as a coordinate on moduli space. In fact, for $\alpha < -1$ the field is not a monotonically decreasing function. This leads to the appearance of three zeros  for a certain range of $x_0$. Hence, we explicitly see the problem existing in the previously discussed self-dual impurity models, where the generalised translation was known only in an infinitesimal form. There, the moduli space metric was calculated numerically, and the zero of the field was used as a moduli space coordinate because of its numerical simplicity. As a consequence, for too negative $\alpha$, this coordinate on moduli space was not globally defined. 

It is important to observe that the Bogomolny equation (\ref{BOG-y}) possesses an additional solution
\be
\phi(y)=\frac{1}{\phi(y)_{BPS}} = -\coth (y+y_0) \label{coth}
\ee
whose existence is related to another symmetry of the Bogomolny equation of the pure $\phi^4$ theory, namely, $\phi(y) \rightarrow 1/\phi (y)$.  
Obviously, this solution is not a regular solution on the full $\mathbb{R}$ due to a singularity at $y=-y_0$. Accordingly, it is not relevant for a construction of the antikink in the presence of the localized impurity. However, as we will see in sec. \ref{annihilation}, this class of solutions of the Bogomolny equation is very important for BPS solutions with vanishing total topological charge. We remark that the $\coth$ solution has been considered in the $\phi^4$ model with a boundary \cite{romanczukiewicz}.
\section{Metric on moduli space and geodesic flow}
Having solved the BPS (self-dual) sector completely, we now analyze some aspects of its dynamics, again in an analytical way. Note that both static and dynamical properties of the non-BPS solitons drastically differ from their BPS counterparts. This is obviously related to the fact that there is no moduli space for non-BPS static solitons. Here we are interested in an analytical description of the BPS sector and do not consider non-BPS solutions. The full Lagrangian is
 \be
 L=\int dx \left[\frac{1}{2}\phi_t^2- \frac{1}{2} \left( \phi_x +\sqrt{2} W + \sqrt{2} \sigma W \right)^2 \right] - \int dx  \sqrt{2} W \phi_x
 \ee
where the topological (boundary) term has been subtracted. Then, the dynamical Euler-Lagrange equation reads
\be
\phi_{tt}-\phi_{xx}- \sqrt{2}\sigma_x W +2WW_\phi (1+\sigma)^2=0.
\ee
In a low-velocity collision of the BPS soliton on the impurity, the system undergoes a sequence of BPS states, which leads to a domain wall that does not get stuck on the impurity \cite{BPS-imp-1}, \cite{BPS-imp-DW}. Hence, it generates a flow on 
moduli space \cite{SM}. To find an analytical form of this flow, we assume that the position of the soliton on the moduli space is a dynamical, time dependent quantity $x_0(t)$, i.e., 
\be
\phi(x,t)=\phi_{BPS}(x, x_0(t)),
\ee
and insert it into the Lagrangian. As a result, we get an effective model which has a nontrivial kinetic part only. The potential part reduces to a 
constant due to the BPS nature of our impurity theory. Here,
\be
L_{eff}= \frac{1}{2} M(x_0) \dot{x}_0^2
\ee
where the effective mass (metric on moduli space) is
\be
M=\int_{-\infty}^\infty dx \left(  \frac{d}{dx_0}  \phi_{BPS} ( x +\Delta_\alpha(x)+x_0(t)) \right)^2 .
\ee
The obvious solution of the effective problem is $$\dot{x}_0(t)=v_{in} \sqrt{\frac{M(x_0=-\infty)}{M(x_0)}} $$where $v_{in}$ is the velocity
of the incoming antikink at minus infinity, which we assume to be the initial state. Note that, since the solution exists for any $x_0  \in \mathbb{R}$, $M(x_0)$ is a positive definite function.  Hence, the metric is globally well defined on the whole moduli space. 
\begin{figure}
\hspace*{0.7cm}
\includegraphics[height=7.0cm]{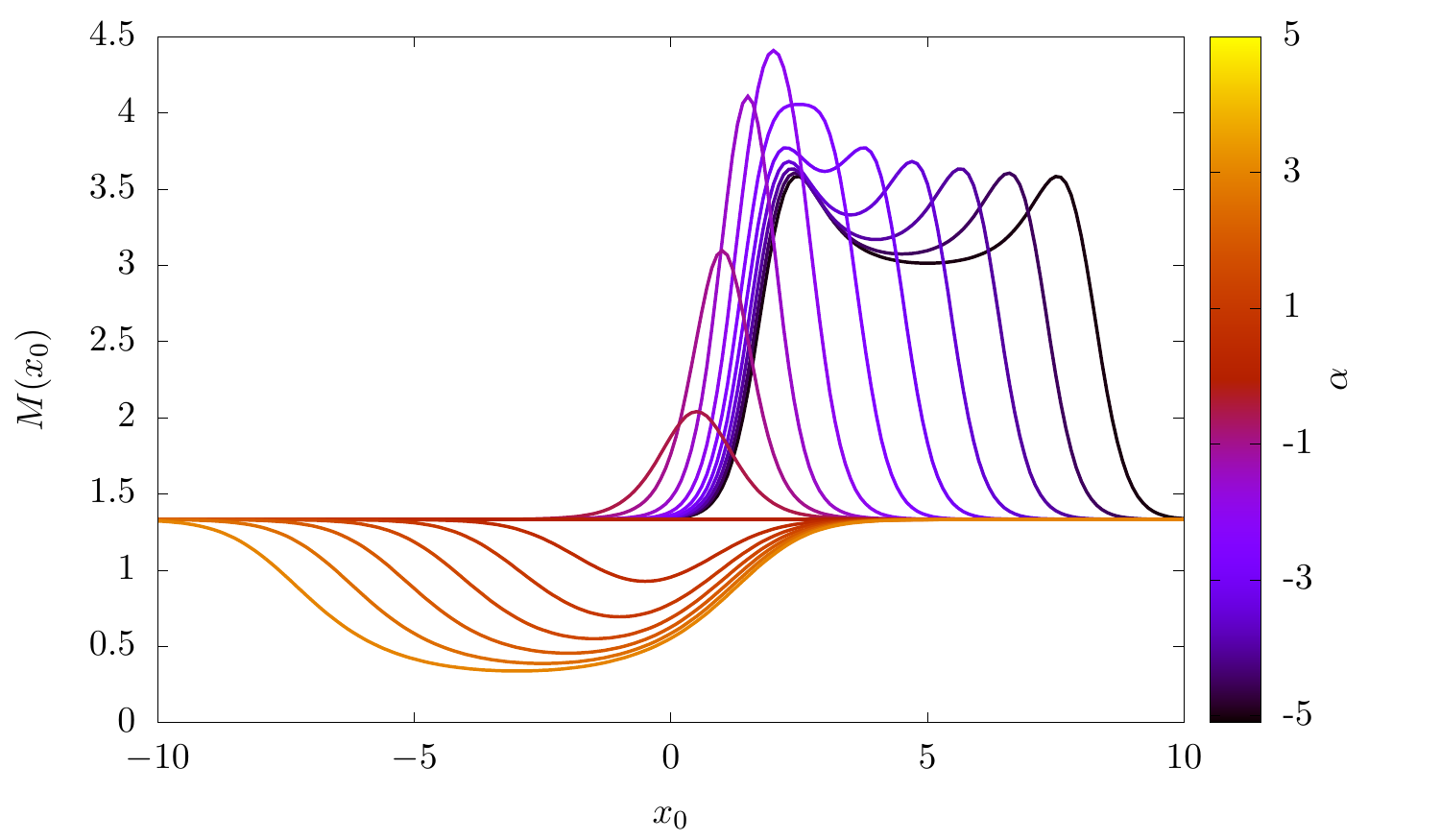}
\caption{Metric on the moduli space for $\phi^4$ BPS-impurity model (\ref{phi4-metric}) for different values of the impurity strength $\alpha$.}
\label{metric-plot}
\end{figure}

For $\phi^4$ theory with the exponentially localized impurity, the metric is given by the expression
\be
M(x_0)= \int_{-\infty}^\infty \frac{dx}{\cosh^4 (x + \alpha \left( \tanh x +1 \right)+x_0(t))} . \label{phi4-metric}
\ee
The integral is finite, greater than 0 for any value of $\alpha$ and any $x_0\in \mathbb{R}$. We plot it in Fig. \ref{metric-plot} for several values of the strength  $\alpha$ of the impurity. For $\alpha>0$ the metric is always below the asymptotic mass, approaching arbitrarily small positive numbers if $\alpha \rightarrow \infty$. For negative impurities, $\alpha \in (\alpha_{crit},0)$ the metric is above the asymptotic mass with one local extremum. 
However, if the strength of the impurity is below $\alpha_{crit}$, three local extrema show up. This structure is related to the particular shape of the impurity rather than to a generic feature of the model like, for example, the existence of three zeros in the profile for some values of $x_0$ if $\alpha < -1$. 

Indeed, e.g., for a rectangular impurity $\sigma(x)=\alpha (\theta(x+1) - \theta (x-1))$ \cite{imp-barrier-1}-\cite{imp-barrier-6}, where $\theta (x)$ is the Heaviside step function, the metric can be derived in a closed form. The shift function is $\Delta_\alpha(x)= \alpha \left( (x+1) \theta(x+1) - (x-1)\theta (x-1) \right)$ which seems to be a monotonously growing function for $\alpha < -1$ and, therefore, the field possesses three topological zeros for some values of the moduli space parameter. Now, the metric reads 
\bea
M(x_0)=\frac{1}{3} \left[ 4\right. &+&\frac{\alpha}{1+\alpha} \frac{\tanh(x_0-1)}{\cosh^2 (x_0-1)} (2\cosh^2(x_0-1)- 1) \nonumber \\  &-& \left. \frac{\alpha}{1+\alpha} \frac{\tanh(x_0+1+2\alpha)}{\cosh^2 (x_0+1+2\alpha)} (2\cosh^2(x_0+1+2\alpha) -1) \right]
\eea
where we assume that $\alpha \neq -1$. 
This metric has no local minima for $\alpha < 0$ but for sufficiently negative $\alpha$ it reveals a plateau. The plateau, for asymptotically large, negative $x_0$, is located at $\frac{8}{3}$.
On the other hand, the BPS antikink solutions possess a very similar structure as in the case of the exponential impurity. Namely, there are three topological zeros for some values of $x_0$, for any $\alpha < -1$. 

\vspace*{0.2cm}

\section{Spectral problem}
The next step of the analysis of dynamical properties of the model, which goes beyond the geodesic approximation, considers small perturbations about the BPS solutions. As the BPS solutions are exactly known, we have the unique opportunity to get an analytical insight into the structure of the oscillating modes as a function of the position on the moduli space.
For this purpose we consider a small perturbation of the BPS solution, $\phi(x,t)=\phi_{BPS}(x) + \cos (\omega t) \eta (x)$. Then, the mode (Schr\"odinger) equation is
\be
-\eta_{xx}+V(x) \eta =\omega^2\eta
\ee
where the effective potential reads
\be
V(x)=2(1+\sigma)^2(W_\phi^2+WW_{\phi \phi}  )- \sqrt{2}\sigma_x W_\phi ,
\ee
and $W, W_\phi, W_{\phi \phi}$ are evaluated for the BPS solution $\phi_{BPS}$. This can also be written as
\be
V(x)=2(1+\sigma)^2 \partial_\phi (WW_\phi) - \sqrt{2} \sigma_x W_\phi=(1+\sigma)^2 \partial^2_{\phi \phi} W^2 - \sqrt{2} \sigma_x \partial_\phi W.
\ee
As the impurity is spatially well-localized at the origin, asymptotically 
(i.e., for $x_0 
\rightarrow \pm \infty$) the BSP antikink solution consists of an infinitely separated free soliton and a lump. Therefore, the  
corresponding spectral problem also decouples. As a consequence, the spectrum of bound states is a 
sum of bound states of the free antikink (in the original no-impurity model) and the lump. Observe that, 
although the lump solutions are spatially constant, they have a nontrivial effective potential in the 
spectral problem. Namely,
\be
V_{\rm lump}(x)= 2\left(1+\sigma\right)^2 W^2_{\phi_{\rm lump}} - \sqrt{2}\sigma_x W_{\phi_{\rm lump}}
\ee
where $W_{\phi_{\rm lump}}$ denotes $W_\phi$ computed on a lump solution, which coincides with one of the vacua of the original, non-impurity model. Thus, $W_{\phi_{\rm lump}}$ is just a model dependent numerical constant. Interestingly, the potential in the spectral problem of the lump solutions takes the form of a super-symmetric quantum mechanical potential 
\be
V_{\rm lump}(x)= U^2(x) - \frac{dU(x)}{dx} 
\ee
where $U(x)=\sqrt{2} W_{\phi_{\rm lump}} (1+\sigma(x))$. Let us assume that we have two symmetric vacua $\phi^\pm$ (as in $\phi^4$ theory). Then, $V^\pm_{\rm lump}=U^2\pm U_x$ and $U=  \sqrt{2} |W_{\phi_{\rm lump}} | (1+\sigma)$. We can use the standard techniques of supersymmetric quantum mechanics to obtain some information about the
spectral problem \cite{dunne}. First, we define two first order differential operators 
\be
A^+=- \frac{d}{dx} +U, \;\;\; A^-=\frac{d}{dx} +U.
\ee
The mode equation can be written as
\be
H_\mp \eta=  A^\pm A^\mp \eta (x) = \omega^2 \eta
\ee
for $\phi^\mp$ lumps. $H_\pm$ are known as partner Hamiltonians. From the last formula it follows that the (possible) zero mode for the $\phi^\mp$ lumps verify the first order equations
\be
A^\mp \eta^\mp_0=0.
\ee
We have two possibilities of the zero mode structure for $\phi^\pm$ lumps depending of the form of $\sigma$. Namely, no normalizable zero modes or only one zero mode (for one of the lump solutions but never for both). Another property of the partner Hamiltonians is that, regardless of the form of $\sigma$, if there are excited bound states, the spectra of both lumps coincide (except for the zero mode). It is also interesting to note that, in the context of SUSY QM, $H_-$ and $H_+$ are the bosonic and fermionic hamiltonians respectively. Therefore, the modes for the $\phi^-$ lump correspond to bosonic states in SUSY QM and the modes for the $\phi^+$ lump correspond to fermionic states in SUSY QM. So we can define a Witten index in our system.

The zero mode equations can be easily solved 
\be
\eta_0^\pm = N e^{\pm \int dx U(x)} = N e^{\pm 2(x+\int \sigma(x)dx)}
\ee
Normalizability of the zero modes requires that the integral $\int \sigma dx $ at infinity grows at least like $x$. But this implies that $\sigma$ at infinity must be a nonzero constant, in contradiction with our assumption of localization. As a consequence, in the background of a lump there are no zero modes for localized impurities. 

\vspace*{0.2cm}

When the BPS antikink moves on the moduli space, approaching the impurity ($x_0 \rightarrow 0$), the spectral structure changes due to a nontrivial deformation of the solution. In some sense, we observe a nonlinear superposition of the modes of the asymptotic states. In particular, oscillating modes (originating both in the free soliton and the lump) move and can even be pushed into the continuous spectrum. This leads to the recently discovered {\it spectral wall} effect \cite{BPS-imp-SW}. 

The simplest version of such a deformation of a mode while moving on the moduli space can be studied 
in the case of the zero mode. In fact, this mode, which is related to the generalized translation, remains a zero mode on the whole moduli space and is given by $
\eta_0= \frac{1}{1+\sigma}\frac{d\phi_{BPS}(x)}{dx}$. Since the BPS solutions are explicitly known, we 
can find the zero mode in a closed form. Moreover, for spatially well-localized impurities, we can conclude that asymptotically the zero mode is confined to the asymptotic (free) soliton. Hence, the zero mode in this asymptotical regime acts nontrivially only on the outgoing/incoming soliton. 

\vspace*{0.2cm}

For example, in the case of the $\phi^4$ model and for an arbitrary impurity, the effective potential takes the following form
\be
V(x)=2\left(1+\sigma\right)^2 (-1+3 \tanh^2(x +\Delta_\sigma(x)+x_0) ) + 2\sigma_x \tanh (x + \Delta_\sigma(x)+x_0 ).
\ee
\begin{figure}
\hspace*{1.2cm}
\includegraphics[height=7.0cm]{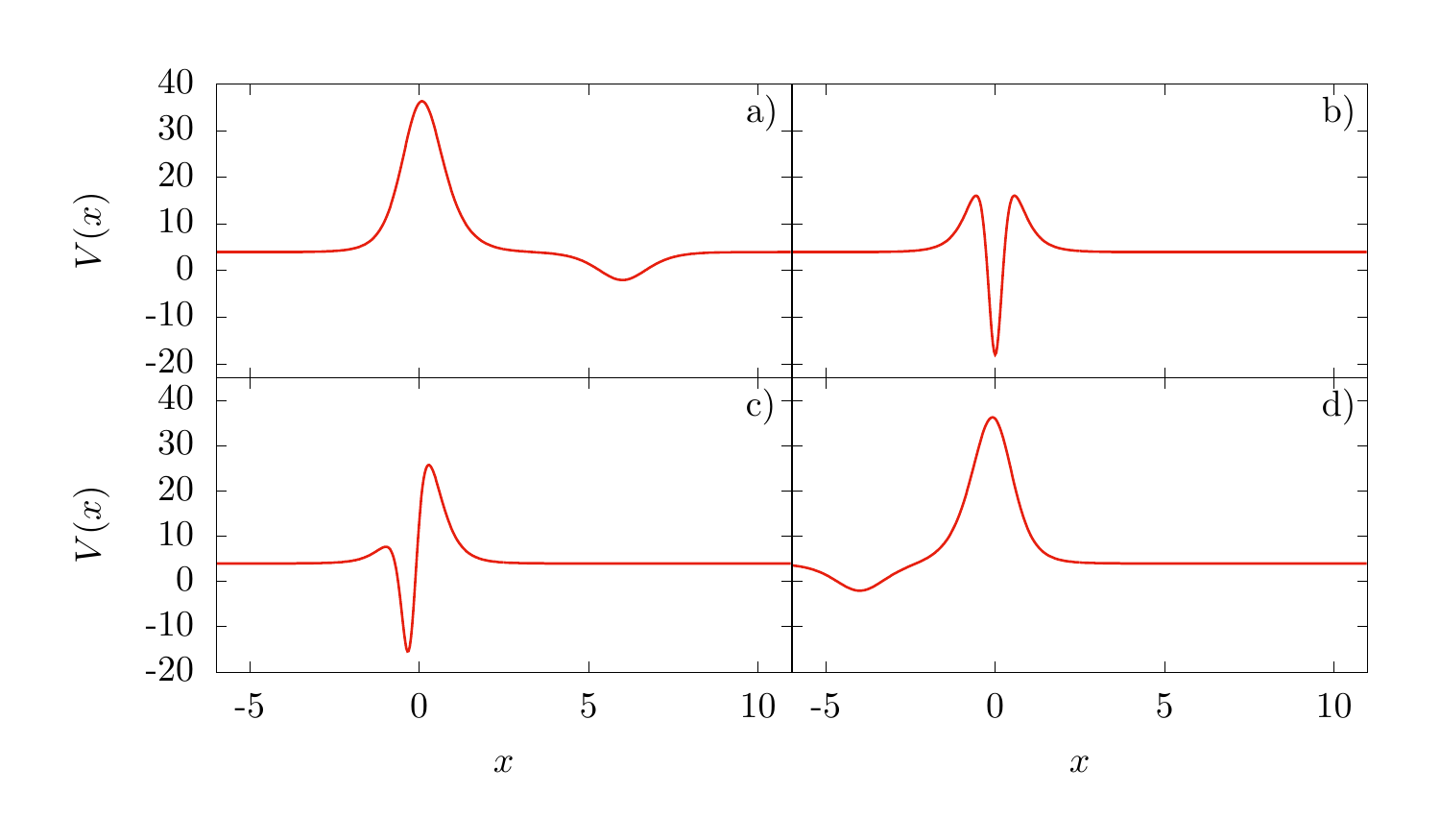}
\\
\hspace*{1.2cm}
\includegraphics[height=7.0cm]{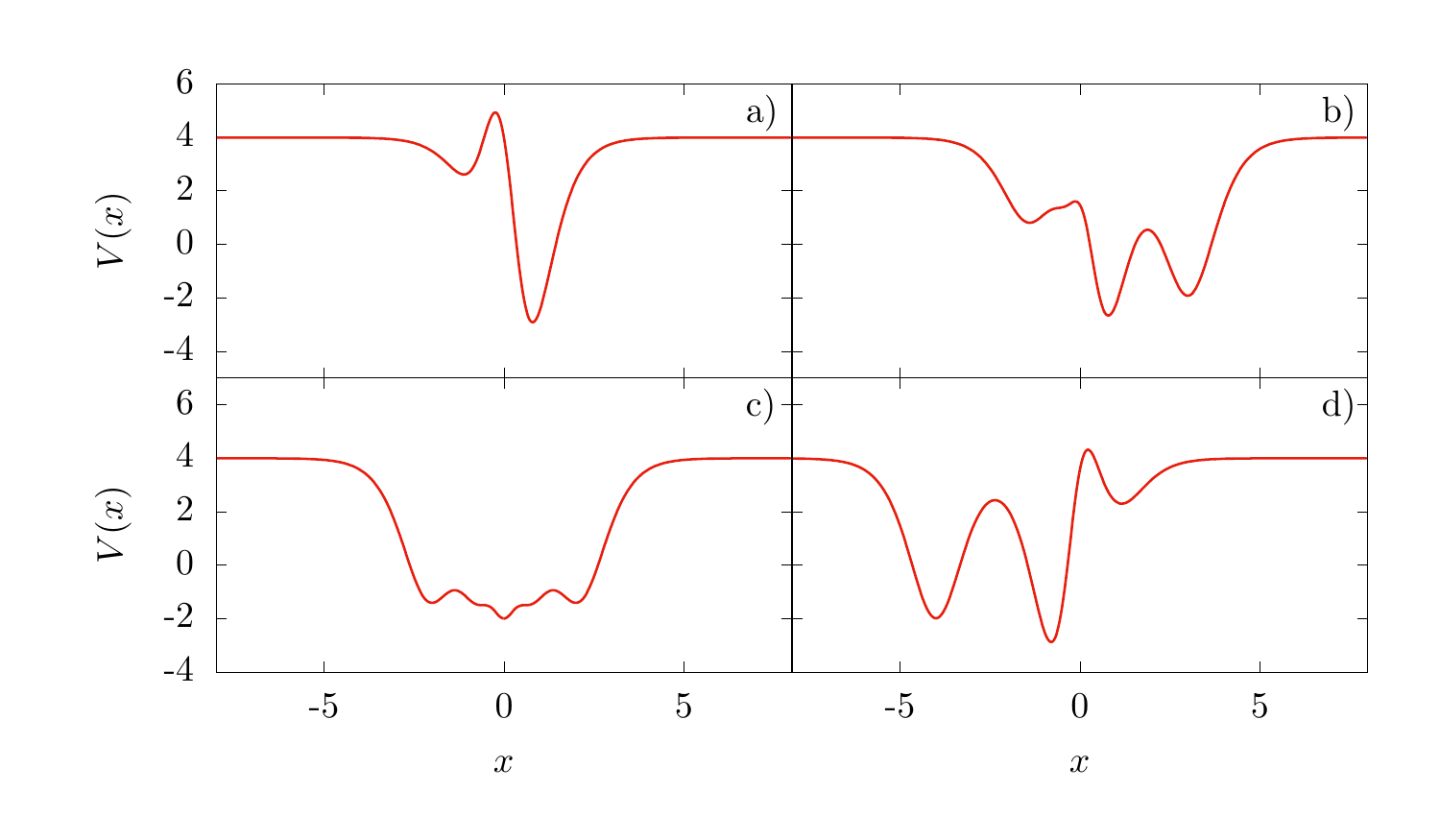}
\caption{Potential $V(x)$ in the spectral problem for $\phi^4$ BPS-impurity model (\ref{eff-pot-phi4}). {\it Left}:  $\alpha=2$ and $x_0=-10, -2, -1, 4$. {\it Right}: $\alpha=-2$ and $x_0=-10, 1,2,4$.}
\label{potential-plot}
\end{figure}
\begin{figure}
\hspace*{1.2cm}
\includegraphics[height=7.0cm]{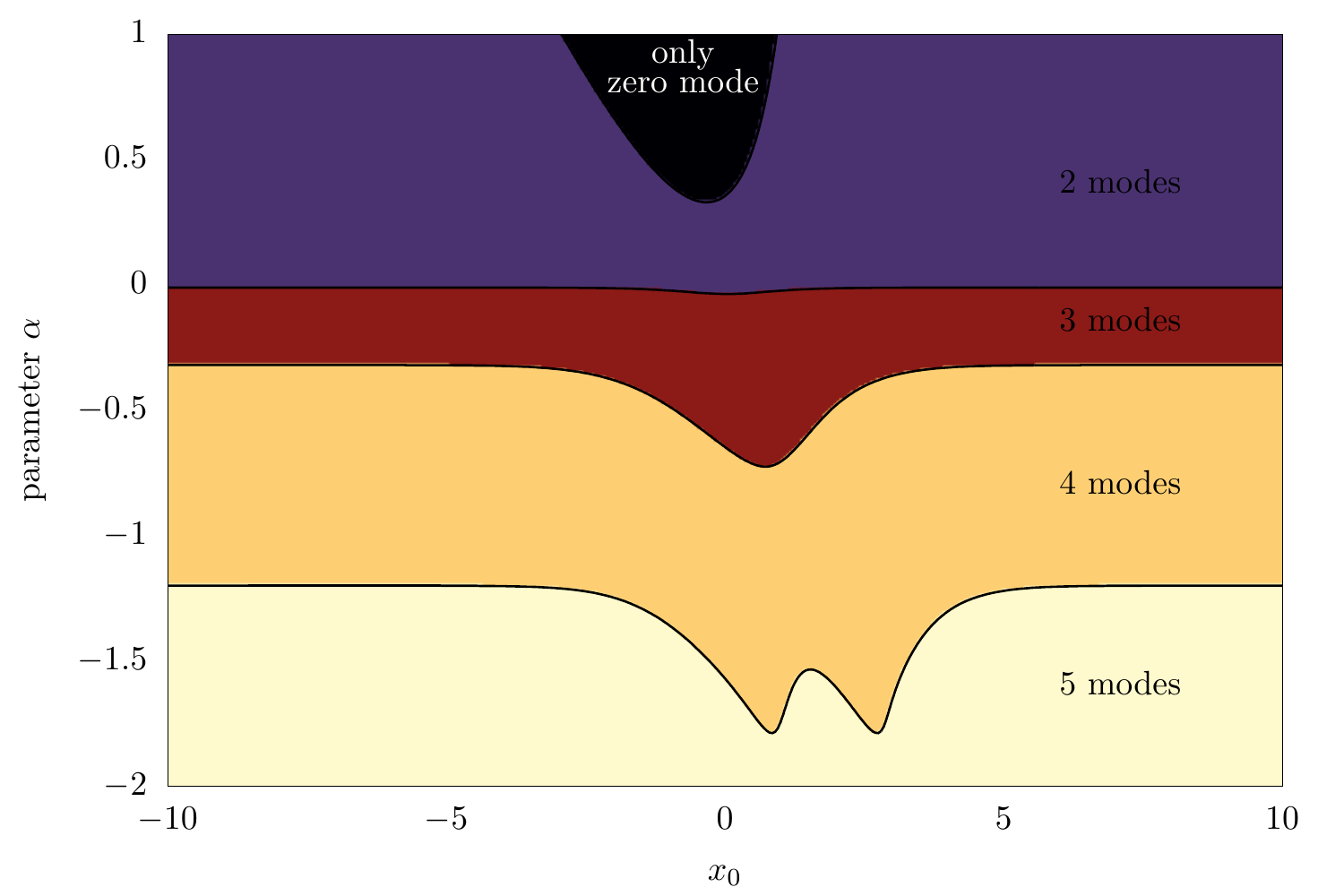}
\caption{Number of discrete modes (zero mode and oscillating modes) in $\phi^4$ BPS-impurity model (\ref{eff-pot-phi4}) in a function of the impurity strength $\alpha$ and the position on the moduli space $x_0$.}
\label{number-modes}
\end{figure}
Now it is perfectly visible that the motion on the moduli space, i.e., the change of the parameter $x_0$, is reflected on the level of the effective potential, and, as a consequence, in the spectral properties of the BPS solutions in a rather nontrivial manner. To be specific, for the exponentially localized impurity we find
\bea
V(x)= 2\left(1+\frac{\alpha}{\cosh^2 x } \right)^2 (&-&1+3 \tanh^2(x + \alpha \left( \tanh x +1 \right)+x_0) ) \nonumber \\ &-& \frac{4 \alpha \tanh x}{\cosh^2 x} \tanh (x + \alpha \left( \tanh x +1 \right)+x_0 ) \label{eff-pot-phi4}
\eea
while
\be
V_{\rm lump}(x)= 4\left(1+\frac{\alpha}{\cosh^2 x } \right)^2 \mp  \frac{4\alpha \tanh x}{\cosh^2 x} 
\ee
where $\mp$ corresponds to the $\phi_{\rm lump}=\pm 1$ lumps, respectively. In Fig. \ref{potential-plot} we show how the effective potential gets deformed when the BPS antikink approaches the impurity. In some sense, we observe a 'collision' of two asymptotically independent spectral problems. In the upper panel (the impurity with $\alpha=2$), in the initial state there is only one oscillating mode localized on the free $\phi^4$ antikink. The impurity does not host any bound state. When the soliton approaches  the impurity, the $\phi^4$ potential well becomes too narrow and shallow, and the mode disappears into the continuous spectrum. In the bottom panel ($\alpha=-2$), the impurity (lump) possesses three oscillating modes. When the BPS soliton meets the impurity, the potential wells merge. In this case, all oscillating modes survive for any position on the moduli space (see Fig. \ref{number-modes}). 

The dependence of the mode structure on the position on the moduli space is strongly sensitive to the details of the assumed impurity (and the field theoretical pre-potential). As is clearly visible in Fig. \ref{number-modes}, the number of modes can be constant during the collision. However, it is not a rare phenomenon that one mode enters the continuous spectrum. It can even reappear again when the soliton is sufficiently close to the impurity. In fact, the richness of possibilities is rather unlimited. Note that a mode which disappears into the continuous spectrum must finally return to the discrete spectrum,  because of the spatial localization of the impurity. 

Whenever a bound mode enters the continuous spectrum, a spectral wall is expected. The family of  solvable self-dual impurity models considered here may, therefore, be ideal for further studies of this phenomenon in a very clear numerical and mathematical set-up. Note that for spatially localised impurities, there always exists a pair of spectral walls, i.e., on both sides of the impurity. 
\section{Non-localized impurities - Step-function examples}
It is another advantage of the present model that it allows for impurities which do not approach zero at infinity. Thus, it does not have to be
 a localized, $L^2$ integrable impurity. This results in a divergent shift function $\Delta_\sigma(x)$. This case is possible 
because of the fact that the impurity is always 
multiplied by the prepotential $W$, ensuring the disappearance of this term if the field takes its vacuum values  (the zeros of $W$). Physically, this means that the impurity changes the 
asymptotic value of the $W^2(1+\sigma)^2$ term, i.e., the asymptotic mass of the soliton.  For instance, let us consider a very simple impurity in the following form, $\sigma = \alpha \theta (x)$. Then, 
\be
\Delta_\alpha = \alpha x \theta (x) \label{theta}
\ee
where $\theta(x)$ is the Heaviside step function. Note that $\sigma$ may be even divergent at infinity. Here, as an example one can consider $\sigma=\alpha x^2$ which leads to $\Delta_\sigma=\alpha x^3/3$. 

Interestingly, such a non-symmetric impurity (\ref{theta}), which can affect the asymptotic mass of the field (at infinity), can compensate a non-symmetric (pre)potential. Here an example may be given by $\phi^6$ theory, where $W=\phi(1-\phi^2)/\sqrt{2}$. In the no-impurity limit the mass of the field (small perturbations around a vacuum) is different at $\phi=0$ ($m=1$) and $\phi=\pm 1$ ($m^2=4$). After coupling the model to an impurity in the self-dual manner, there are two BPS solutions: the antikink $1\rightarrow 0$
\be
\phi_{\bar{K}_{BPS}}=\sqrt{\frac{1-\tanh (x+x_0+\Delta_\sigma(x))}{2}}
\ee
and the kink $-1 \rightarrow 0$
\be
\phi_{K_{BPS}}=-\sqrt{\frac{1-\tanh (x+x_0+\Delta_\sigma(x))}{2}} .
\ee
The two other static solitons are not BPS solutions. Then, for the BPS antikink and kink the mass of small perturbations at $\phi=+1$ and at $\phi=0$ are equal if $\Delta=\theta (x)$. For the non-BPS solitons, the masses of the field at plus and minus infinity are different. This may affect the resonant, fractal-like structures in a kink-antikink collision.
\begin{figure}
\hspace*{2.0cm}
\includegraphics[height=6.5cm]{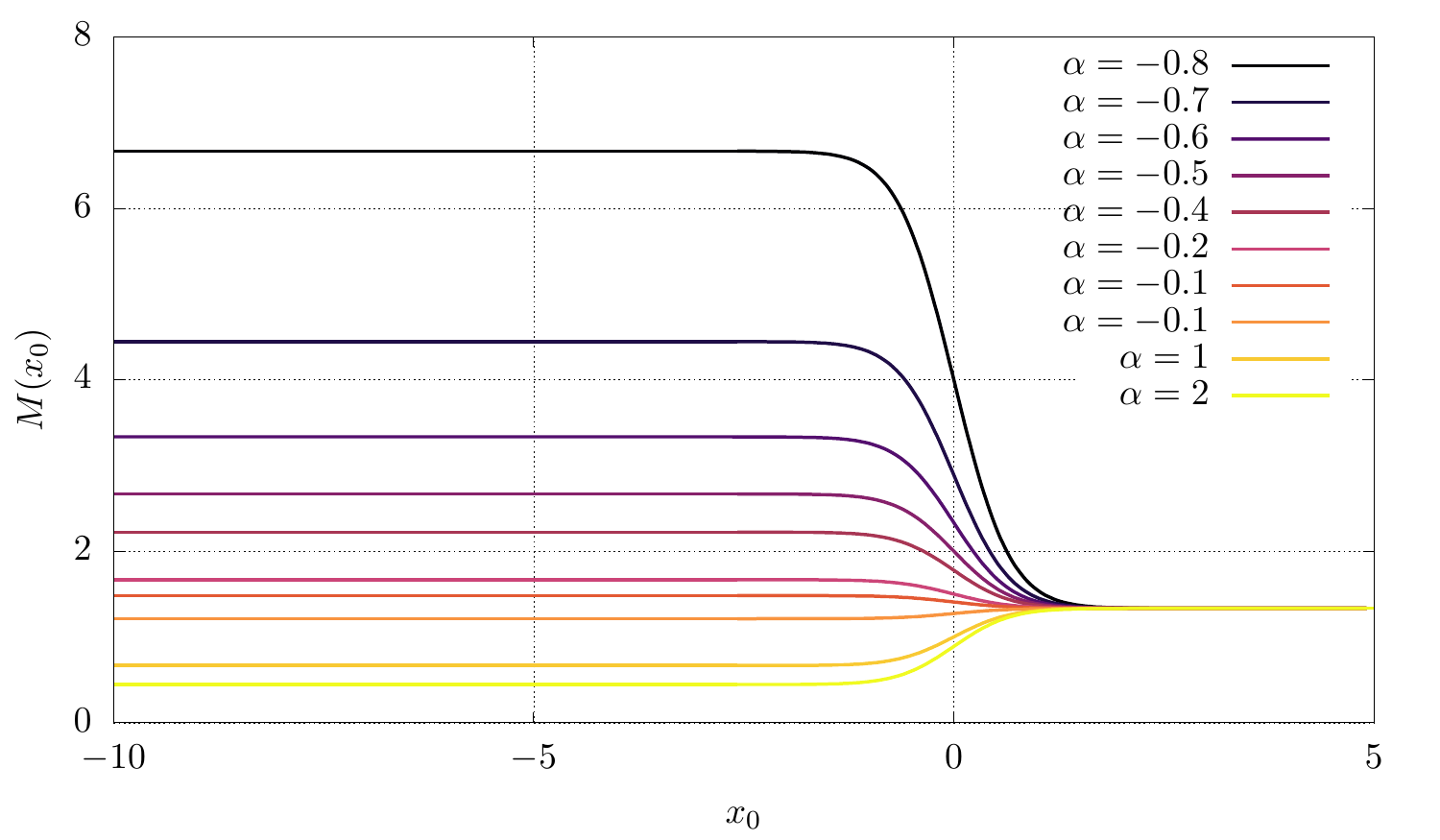}
\caption{Metric on the moduli space for $\phi^4$ BPS-impurity model with non-localized, step function impurity (\ref{step-metric}) for different values of the impurity strength $\alpha$.}
\label{metric-step-plot}
\end{figure}

If we assume the step function impurity \cite{step-imp}, then the metric can be computed in an exact form. For the $\phi^4$ model, it reads
\be
M(x_0)= \frac{\alpha}{1+\alpha} \left( \tanh x_0 -\frac{1}{3} \tanh^3 x_0 \right) +\frac{2}{3} \left( 1+\frac{1}{1+\alpha} \right) \label{step-metric}
\ee
for $\alpha >-1$. As remarked before, the asymptotic masses of the solitons at plus and minus infinity (on the moduli space) are not the same. Namely, $M(x_0=\infty)=\frac{4}{3}$ and $M(x_0=-\infty)=\frac{4}{3}\frac{1}{1+\alpha}$. One can check that for any $\alpha > -1$ the metric is globally well-defined on the whole moduli space i.e.,  $M(x_0)>0$ for all $x_0 \in \mathbb{R}$. For $\alpha>0$ $(\alpha <0)$ the metric function is a strictly growing (decreasing) function, see Fig. \ref{metric-step-plot}. 
Similarly, for $\phi^6$ BPS solitons and the step function impurity we get
\be
M(x_0)=\frac{1}{32(1+\alpha)} \left(4+\alpha +2\alpha \tanh x_0+\alpha \tanh^2 x_0 \right) 
\ee
which monotonically interpolates between $M(x_0=\infty)=\frac{1}{8}$ and $M(x_0=-\infty)=\frac{1}{8(1+\alpha)}.$

Here, as the effective potential does not have to tend to the same asymptotic value at $\pm \infty$, the spectral wall may exist only on one side of the impurity, without any partner.

\section{Self-dual kink-antikink solutions}
\label{annihilation}
\subsection{Annihilation of solitons on the moduli space}
There is a class of non-localized impurities for which the Bogomolny equation supports 
topologically trivial solutions only. 
Surprisingly, such solutions result in a moduli space with a nontrivial metric. Consider, for example,
 $\phi^4$ theory with the following impurity
\be
\sigma_j (x)=\frac{j}{2} \tanh x -1
\ee
where $j$ is a positive integer number.  The shift function is now $\Delta_j (x)= \frac{j}{2} \ln \cosh x -x$, 
which leads to the following solution of the Bogomolny equation
\be
\phi (x)= - \tanh \left( \frac{j}{2} \ln \cosh x +x_0 \right) = -\frac{\cosh^j x - a}{\cosh^j x + a} \label{sol-triv}
\ee
where $a=e^{-2x_0}$.
In spite of the fact that we used a formal antikink BPS solution in the $y$ coordinate (treated as a solution of (\ref{BOG-y})), this solution is 
topologically trivial. Indeed, the BPS solution (\ref{sol-triv}) has a 
maximum for $x=0$, where it reaches $-\tanh x_0$, and then it approaches the  same value at $\pm\infty$, i.e., $\phi(x = \pm \infty)=-1$. Treated as an element of the moduli space, the BPS solution for $x_0 \rightarrow \infty$ ($a\rightarrow 0$) approaches
the lump solution $\phi_{lump}^-=-1$. When $x_0$ decreases, the solutions develop a hill centered at $x=0$. 
By further decreasing $x_0$, the hill grows until it develops a plateau at $\phi \rightarrow +1$. For 
$x_0 \rightarrow -\infty$ ($a\rightarrow \infty$), the plateau expands and the solution represents a kink-antikink 
pair with the separation distance growing to infinity. In other words, the motion on moduli space, 
from $x_0=-\infty$ to $x_0=\infty$, is equivalent to the scattering of an initially infinitely separated pair of a kink and an antikink. At the final stage, the solitons completely annihilate and we have the lump solution $\phi^-_{lump}$ only. 
Observe that, as all BPS solutions have the same energy, there is no static force between the 
constituents, kink and antikink. However, this three body (kink-impurity-antikink) system is not completely free, as the solitons must be at equal distance from the impurity. 

\begin{figure}
\hspace*{1.5cm}
\includegraphics[height=6.5cm]{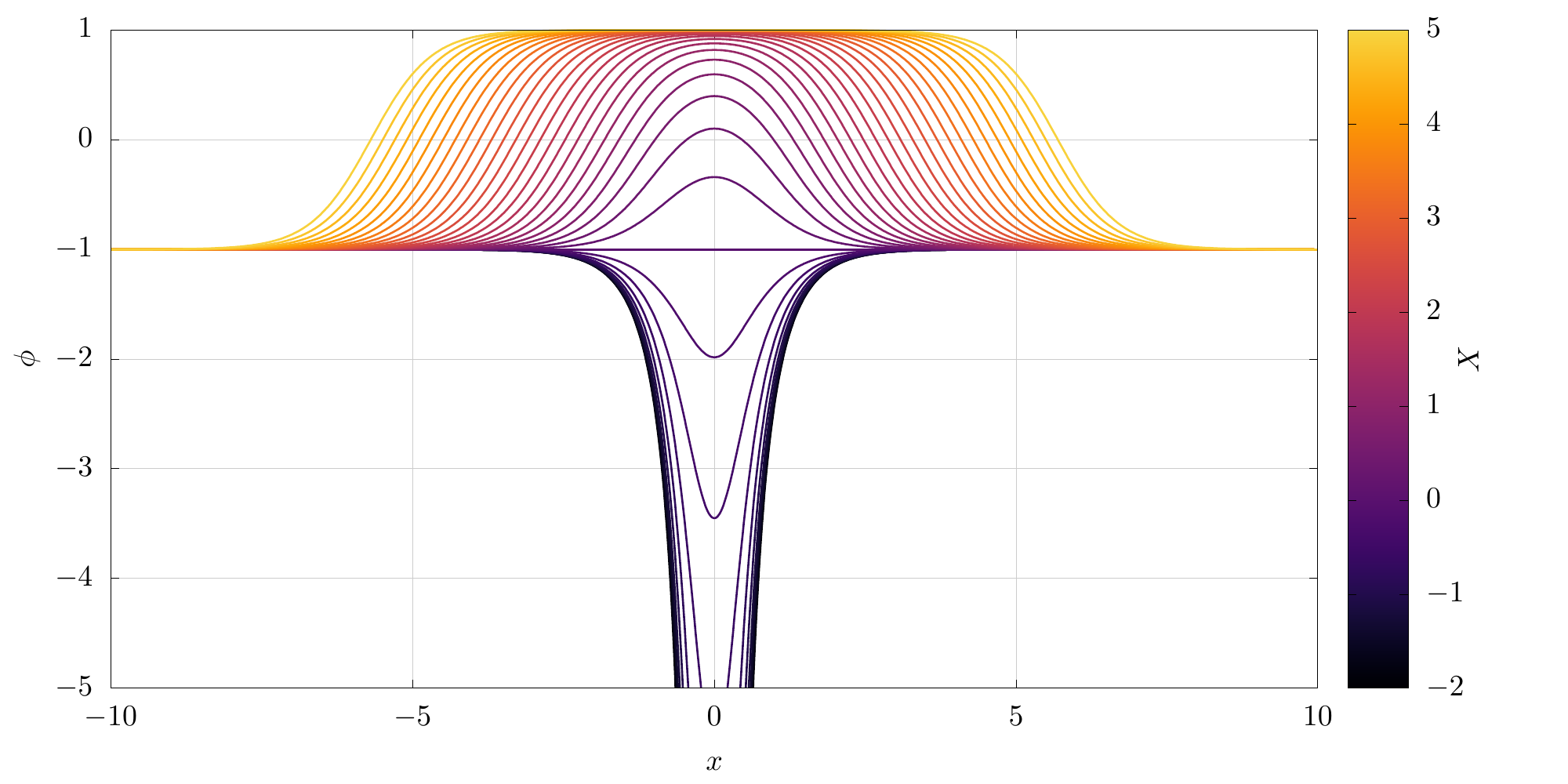}
\caption{BPS solutions for different values of the moduli parameter $X$ for the annihilating impurity with $j=1$.}
\label{scatter-sol}
\end{figure}

Surprisingly, it turns out that $x_0$ does not cover the whole moduli space for this impurity, and that the full moduli space of topologically trivial solutions is even larger. Indeed, one can easily verify that the BPS solution (\ref{sol-triv}) remains a solution of the Bogomolny equation if the parameter $a$ takes negative values. If $a \leq -1$ the solution diverges at $x_{sing}$ such that $\cosh^jx_{sing}+a=0$. Therefore, these are not physically acceptable solutions. On the other hand, for $a \in (-1,0)$, the solutions are globally well-defined solutions of the Bogomolny equation and, as a consequence, must contribute to the moduli space. We plot this new, negative part of the BPS solutions (together with the first, positive part) in Fig. \ref{scatter-sol} in a new moduli space coordinate $X \in (-\infty, \infty)$, defined as 
\be 
a=-1+e^{jX}.
\ee 
This new moduli space coordinate better describes the zero of the field, i.e., the position of the soliton in the large separation limit. However, in some formulas we will still keep $a$ due to its simplicity. For $a=0$ ($X=0$) we start with the $\phi=-1$ lump, which was the end of the first, positive $a$ family of the BPS solutions. Once $a$ (or $X$) takes negative values, the field develops a dip located at $x=0$. As $a$ tends  to -1 ($X$ tends to $-\infty$) the dip approaches an arbitrary large (negative) value. 

It is interesting to notice that, while the positive part of the BPS solutions emerges from a {\it regular} solution of the original (no-impurity) $\phi^4$ theory, the negative part stems from the {\it singular} solutions of the no-impurity model.  Indeed, the Bogomolny equation (\ref{BOG-y}) possesses the following solution
\be
\phi(y)=-\coth y
\ee
which is singular at $y=0$, see our discussion below eq. (\ref{coth}). 
However, in the case of the impurity $\sigma_j$, the variable $y$ is expressed via $x$ as  $y=\frac{j}{2}\ln \cosh x +\tilde{x}_0$, where $\tilde{x}_0 $ is a constant.  Hence, $y$ is bounded from below by $y_{min}=\tilde{x}_0$ and the solution does not reach the singularity if $\tilde{x}_0 >0$. As a consequence, we get 
\be
\phi(x)=-\coth \left(  \frac{j}{2} \ln \cosh x +\tilde{x}_0 \right)=-\frac{\cosh^j x +e^{-2\tilde{x}_0}}{\cosh^jx-e^{-2\tilde{x}_0}},
\ee
that is, a solution (\ref{sol-triv}) with $a=-e^{-2\tilde{x}_0}$. The fact that the moduli space needs to be complemented by the originally (no impurity theory) divergent solutions is a new and rather unexpected result, which may have some impact on the construction of the moduli space of other self-dual impurity models, also in higher dimensions \cite{BPS-imp-2}. 

To summarize, the full motion on the moduli space describes an annihilation of the initially infinitely separate kink-antikink pair. The process does not stop at the $\phi=-1$ lump but goes through this point and ends on a well formed by solutions corresponding to the singular solutions of the no-impurity model. Note that such a well is observed in kink-antikink collisions in the pure $\phi^4$ theory. In fact, during the collision, the solitons form a sort of well with a shape that may be approximated by 
\be
\phi(x)=-\tanh(x-X)+\tanh(x+X)-1
\ee
with $X>-1$. At the minimum of this well, the field takes the value $-3$, at most. In this regime, our negative type of solutions reproduces this behavior. Therefore, the appearance of the negative type of solutions should be considered as an advantage of the impurity model. Of course, in contrast to the pure $\phi^4$ theory, in our model there is no mechanism preventing the well from developing an arbitrarily large depth.   

\begin{figure}
\hspace*{2.0cm}
\includegraphics[height=6.5cm]{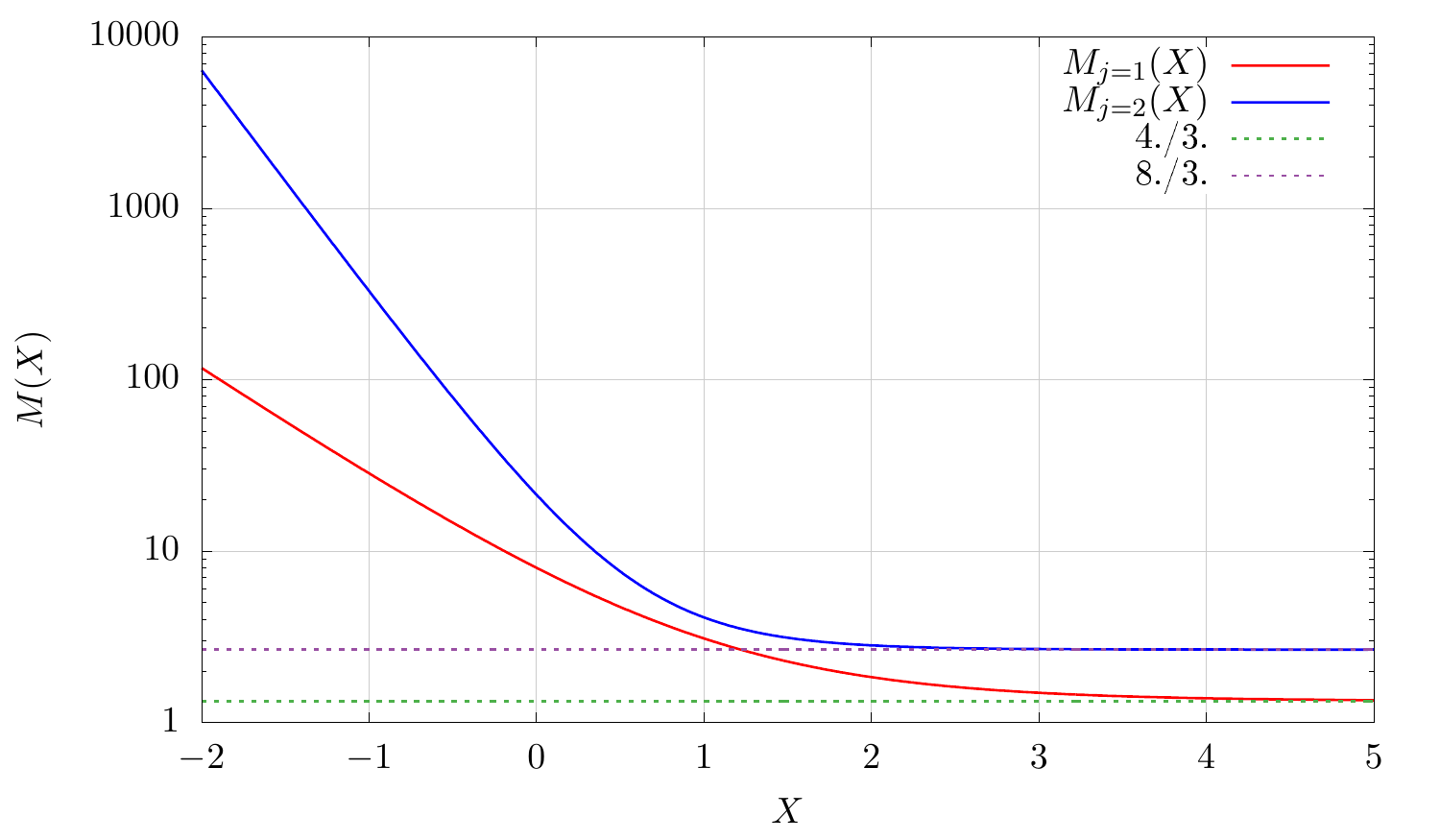}
\caption{Metric on the moduli space for the annihilating impurity with $j=1,2$. The horizontal line denotes the asymptotic values $M(X=\infty)$. A logarithmic scale is used for $M(X)$, because it grows very fast for negative $X$.}
\label{scatter-M}
\end{figure}

\hspace*{0.2cm}

The low speed dynamics of  this annihilation process is described by the geodesic approximation. 
The moduli space metric for the corresponding topologically trivial BPS sector is
\be
M_{j}(a)= 4 \int_{-\infty}^\infty \frac{\cosh^{2j}x dx}{\left(\cosh^j x +a \right)^4} 
\ee
which for $j=1$ has the following exact form
\be
M_{j=1}(a)= -\frac{4}{3(1-a^2)^3} \left( a^4-10a^2-6 - 6a\frac{4+a^2}{\sqrt{1-a^2}} \arctan \frac{a-1}{\sqrt{1-a^2}} \right) 
\ee
In Fig. \ref{scatter-M}, we plot the metric $M(X)$ corresponding to the
collective coordinate $X$, related to $M(a)$ via
$M(X) = (da/dX)^2 M(a) = j^2 (a+1)^2 M(a)$. In both coordinates, the
metric is a monotonic function which smoothly interpolates between infinity at $X=-\infty$ ($a=-1$) and a finite value at $X=\infty$ ($a=\infty$). Specifically, for asymptotically large $a$ the metric tends to zero in a power like manner,  $M_{j=1}(a)= (4/3)a^{-2}+O(a^{-4})$. Thus, the zero of the metric is approached in infinite time $a \propto e^{t \cdot const}$. In $X$ coordinate the metric tends to a nonzero value $M_{j=1}(X=\infty)=4/3$ also in infinity time
\be
X \propto t
\ee
For $a\rightarrow -1$ the metric asymptotically reads $M_{j=1}(a)=\frac{5\pi}{2\sqrt{2}} (1+a)^{-7/2}+O((1+a)^{-5/2})$. Therefore, the infinity of the metric is approached again in infinite time $(1+a) \propto t^{-4/3}$ or
\be
X \propto \ln t
\ee
Finally, the point $X=0$ ($a=0$) where the positive and negative type of solutions are connected is a completely regular point of the the metric. Here $M_{j=1}(a=0)=8$. All this means that the singular branch of BPS solutions (with $a\leq -1$) is dynamically separated from the regular branch (formed by both positive and negative types of solutions smoothly joined at $a=0$). Indeed, the $a\rightarrow -1$ boundary is approached for $t\rightarrow \infty$.

An analogous behavior is found for the other impurities $\sigma_j$. For example, for $j=2$ the moduli space metric is
\be
M_{j=2}(a)=\frac{1}{6} \left( \frac{-3+4a(4+a)}{a(1+a)^3} +\frac{3(1+6a)}{(1+a)^{7/2} a^{3/2}} \mbox{arctanh} \sqrt{\frac{a}{1+a}} \; \right) .
\ee
Again, the metric has a finite value at $a=0$ (or $X=0$), $M_{j=2}(a=0)=16/3$, where the positive part smoothly joins the negative part of the self-dual solutions.

\hspace*{0.2cm}

To some extent, the obtained moduli space metric reveals some similarities with the metric for the $CP^1$ $\sigma$-model in (2+1) dimensions \cite{leese}. In the simplest set-up, for two or more solitons sitting on top of each other, the moduli parameter is given by the height of the soliton $\lambda$, while its position is kept fix. Hence, the complex field which is the primary field of the model reads $u=\lambda^n (x+iy)^n$, where $n$ is the pertinent topological charge carried by the soliton.  When $\lambda$ is small, the soliton is small and broad, while for large $\lambda$ the soliton gets narrow and tall. Then the metric is just $M(\lambda)=2C_n \lambda^{-4}$, where $C_n$ is a constant \cite{cp1-zakrzewski}. The zero of the metric, which corresponds to the soliton collapsing to an infinitely thin and peaked configuration, is approached in a finite time as $\lambda \sim 1/(t_{crit}-t)$. At this point, the configuration possesses a singularity in the energy density, and the geodesic approximation as well as the full numerics break down \cite{cp1-zakrzewski}. In our case the zero as well as the infinity of the metric are approached in infinite time. Hence, we have a globally well defined geodesic dynamic. 

\hspace*{0.2cm}

From a more general perspective, this is a unique example of a solitonic model which has a BPS sector with soliton-antisoliton solutions. Typically,  if a Bogomolny equation supports multisoliton configurations, the constituents carry either positive or negative topological charge, but never both. This is exactly what happens in the Abelian Higgs model at critical coupling, where a BPS solution is provided by a rational function of an either holomorphic or antiholomorphic variable. The same occurs for self-dual instantons, where instanton-antiinstanton solutions do not belong to the BPS sector. Static soliton-antisoliton solutions are rather examples of {\it sphalerons}, i.e., unstable solutions which are not minima but mountain pass solutions of the energy functional. Here, the $CP^N$ model with $N>1$ may serve as an example, see \cite{Zakrzewski}. Static soliton-antisoliton configurations do not obey a first order Bogomolny type equation and do not saturate any topological energy bound and, therefore, are not BPS solutions. 
Observe that a BPS sector containing solitons with positive and negative topological charge (coexisting holomorphic and antiholomorphic maps) has been discovered very recently for other BPS-impurity theories in 2+1 dimensions. Namely, in the BPS-impurity $CP^1$ model \cite{BPS-imp-2} and later in the impurity Abelian Higgs model \cite{han}. Furthermore, mixed holomorphic and antiholomorphic maps are BPS solutions of a model of the so-called magnetic Skyrmions at the critical coupling \cite{schroers-1}-\cite{schroers-2}. In fact, a closed relation between this theory and self-dual impurity models has been recently pointed out in \cite{BPS-imp-2} and \cite{schroers-2}. 

Note also that both the kink and the antikink constituents, if considered separately, are not solutions of the Bogomolny equation and, therefore, are not contained in the BPS sector. Only their bound state, although with zero binding energy, solves the Bogomolny equation. 

We remark that a collective coordinate analysis with the relative distance between kink and antikink as one collective coordinate was applied to the kink-antikink interaction in the usual, no impurity, scalar field theory in (1+1) dimension. However,  such a kink-antikink configuration is not a BPS solution, and the collective coordinate does not belong to a moduli space. It is, therefore, not guaranteed that this coordinate covers the dynamics of this process on the whole $\mathbb{R}$, even approximately. This is one of the reasons why there is no correct analytical description of the annihilation process in, e.g., the $\phi^4$ model \cite{weigel-1}-\cite{weigel-2}. 
\subsection{Annihilation and the spectral structure }
Another important observation is that this choice of the impurity leads to the P\"oschl-Teller potential in the spectral problem of the lump solutions. Indeed, we find that $U=j \tanh x$ and 
\be
V^\pm_{\rm lump} = j^2 - j(j\mp 1)\frac{1}{\cosh^2 x}.
\ee
Hence, we trivially find the spectral structure of the lump solutions. For $\phi_{\rm lump}=-1$ (which does 
belong to the moduli space of the BPS solution) the potential is $V_{\rm lump}^-$. Thus, there are $j$ 
discrete bound states with energies $E_n=n(2j-n)$, $n=0,1,...,j-1$, which includes a normalizable zero 
mode $\eta^-_0=(\cosh x)^{-j}$. The $\phi_{\rm lump}=1$ leads to the $V_{\rm lump}^+$ potential, which possesses 
$j-1$ discrete modes with energies coinciding with the negative lump, except, of course, the zero mode 
which is absent. The continuous spectrum starts above $E=j^2$ in both cases. The existence of the zero 
mode for the fixed impurity can be, at first glance, a surprising feature. However, in this set-up the lump 
solution is not an isolated topologically trivial solution but it {\it belongs to the moduli space}. So, it is 
precisely the zero mode responsible for the generalized translation which provides the motion on the moduli space.   

The annihilation process is particularly transparent on the level of the scattering of the potentials in the spectral problem. In fact, the full expression of the spectral potential reads
\be
V(x)=\frac{j^2}{2} \tanh^2 x \left( -1 +3 \left( \frac{\cosh^jx-a}{\cosh^j x +a}  \right)^2  \right)- \frac{j}{\cosh^2 x} \cdot \frac{\cosh^jx-a}{\cosh^j x +a}.
\ee
\begin{figure}
\hspace*{1.0cm}
 \includegraphics[height=10.0cm]{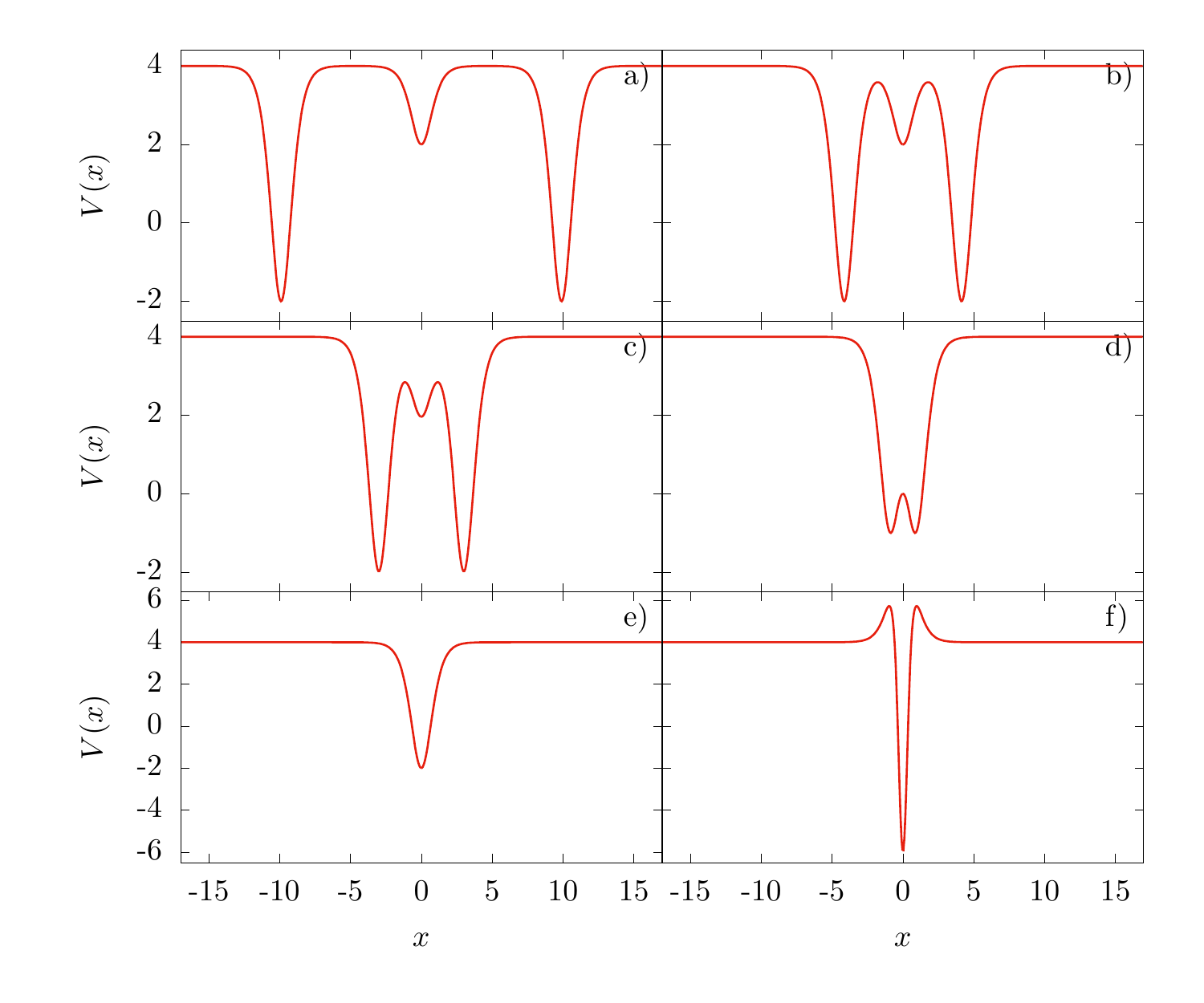}
\caption{The effective potential for the annihilating impurity $j=2$ for $a=10^8, 10^4, 10^2, 1, 0, -0.5.$}
\label{pot-scatter}
\end{figure}
At one boundary of the moduli space, when $a \rightarrow \infty$ ($X\rightarrow \infty$), the potential tends to three separate potentials which are joined smoothly, consisting of the $V^+_{\rm lump}$ at the origin and two $\phi^4$ type (anti)kink potential wells approaching the spatial infinities, see Fig. \ref{pot-scatter} a) where $a=10^8$ is plotted. (Of course, they quantitatively reproduce the $\phi^4$ effective potentials only for $j=2$.) This regime corresponds to the well-separated kink-antikink pair. As $X$ decreases, the soliton potential wells approach the central region and merge, forming the $V_{\rm lump}^-$ for $a=0$ (or $X=0$), Fig. \ref{pot-scatter} e). It is clear that the zero mode, initially confined to the incoming solitons, gets trapped on the lump solution $\phi=-1$. For negative $a$ (or $X$) the effective potential looks like a well located at $x=0$, which becomes deeper and narrower as we approach the second boundary of the moduli space at $a\rightarrow-1$ ($X\rightarrow - \infty$).

\begin{figure}
\hspace*{-1.0cm}
  \includegraphics[height=5.0cm]{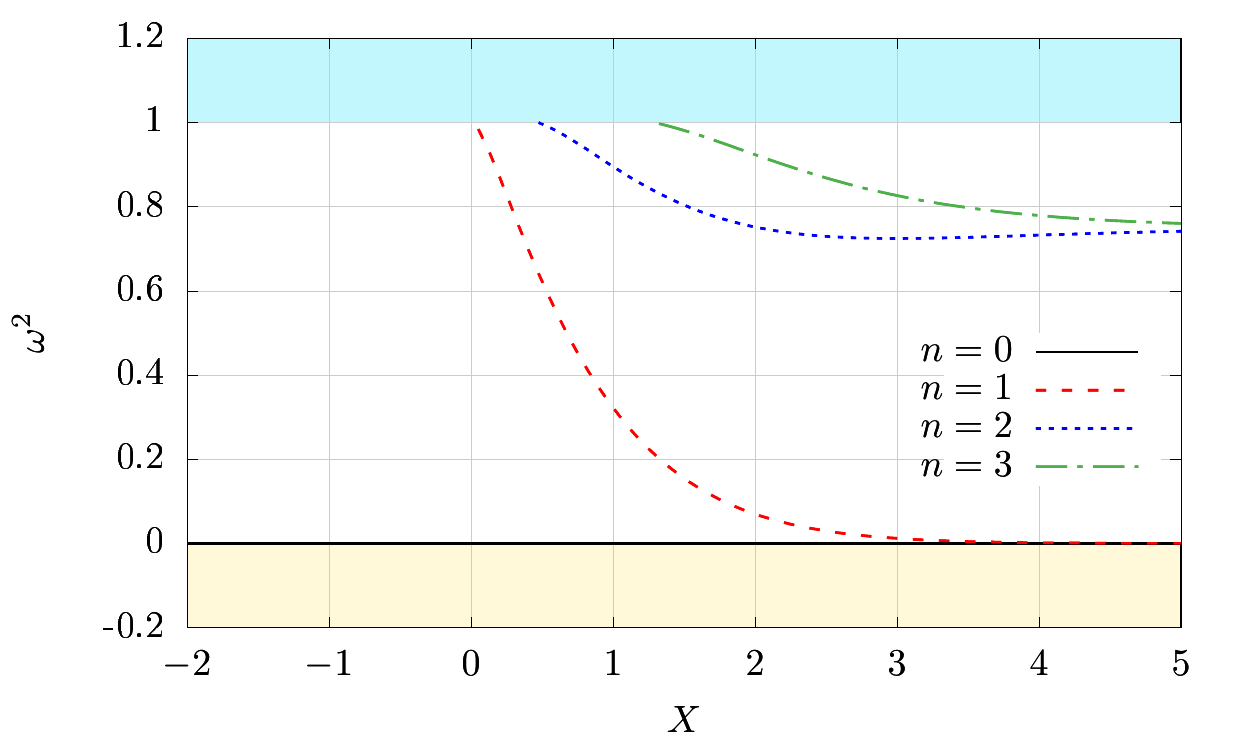}
  \includegraphics[height=5.0cm]{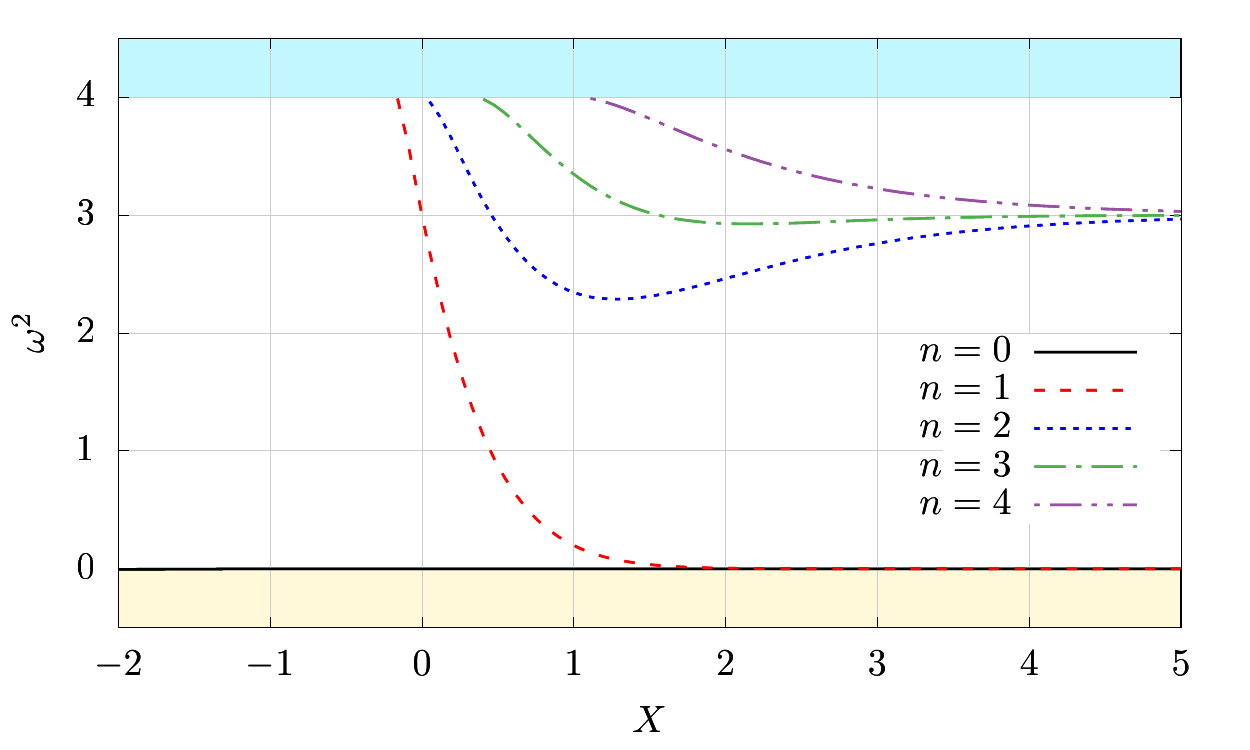}
\caption{Dependence of the spectral structure on the position on the moduli space $X$ for $j=1$ (left) and $j=2$ (right) annihilating impurities.}
\label{modes-scatter}
\end{figure}

Obviously, the observed changes in the form of the potential in the spectral problem are reflected in changes of the mode structure while moving on the moduli space. In the case of the $\sigma_{j=1}$ impurity for $X=\infty$, the mode structure of $\phi^4$ theory is reproduced, up to a multiplicative factor, as visible in Fig. \ref{modes-scatter} left panel. This is because the $\phi=1$ lump solution does not have any bound modes. Hence we have two zero modes (translations of each soliton) and two discrete modes. As $X$ decreases, the zero mode splits into a zero mode and a discrete mode whose frequency rises until it hits the continuous spectrum. The asymptotic bound modes also split and enter the continuous spectrum. Then, for negative $X$, the potential well hosts only the zero mode. In fact, it is a general feature of the spectral problem that as $X$ approaches $-\infty$ all bound modes get expelled into the continuous spectrum. The potential well becomes simply too narrow. Thus, finally, for sufficiently small (negative) $X$ there is only the zero mode. The same pattern occurs for the $\sigma_{j=2}$ impurity, see Fig. \ref{modes-scatter}, right panel. 

Both impurities may be used for a further investigation of the kink-antikink annihilation process in the $\phi^4$ model and the possible creation of an oscillon. Indeed, this model 
gives us the rare opportunity to study the role of internal modes in the annihilation process in a set-up 
where there is no static force between kink and antikink. In the next step, such a static force may be switched on by the addition of an arbitrarily small new potential or impurity term,  which breaks the self-duality explicitly. As a result, we could, in a controlled manner, analyze the mutual impact of the intersoliton 
static force and the excited modes structure on the scattering of solitons in the topologically trivial sector. Such a self-duality breaking term will also prevent the effective potential well to reach  an arbitrary depth.

\vspace*{0.2cm}

Let us underline that the particular form of the impurity assumed here was chosen to provide a 
solvable spectral problem (for the lump solution). The observed topologically trivial moduli space can be 
found for many other examples. The only condition is that the impurity leads to a shift function such that 
$x+\Delta_\sigma (x)$ approaches plus (or minus) infinity for both $x \rightarrow -\infty$ and $x\rightarrow \infty$. We can conjecture that whenever an impurity hosts a zero mode, the Bogomolny equation supports topologically trivial solutions only. As a consequence, there is a moduli space of charge $Q=0$ BPS solutions describing an annihilation/creation process of solitons. 

\section{Summary}

In this article, we have presented a large family of solvable (half) self-dual soliton-impurity models in (1+1) 
dimensions. Like all half self-dual soliton-impurity models, these models possess a BPS (self-dual) 
sector to which half of the static solitons belong. Hence, these solitons saturate the pertinent 
topological energy bound and obey a Bogomolny type equation. 
In comparison with other field theories in (1+1) dimensions,
one exceptional property of these models is that different solutions of the Bogomolny equation vibrate differently. In other words, the spectral structure depends on the position on the moduli space.

The first new finding is that this particular family of theories is completely solvable in the self-dual sector 
for any spatial distribution of the impurity. This leads to exact expressions both for topological BPS solitons and 
lumps, 
i.e., topologically trivial solutions. Furthermore, an exact expression for the finite (i.e., non-infinitesimal) generalised translations was also 
found. This generalised translation is a symmetry of the Bogomolny equation (not the full action) transforming one self-dual 
solution into another, energetically equivalent solution. Hence, it generates the whole moduli space 
of BPS states. As the generalized translation is known in an exact form, we have an analytical 
description of the moduli space together with its metric. This allows for an exact understanding of the 
geodesic flow, i.e., the low energy interaction (scattering) of the incoming BPS soliton on the impurity. 
In addition, we have established a rather involved relation between the position on the moduli space and the
form of the effective potential in the small perturbation problem, which defines the spectral structure 
(vibrational modes) of the solution. 

The second main result is that the self-dual impurity models may lead to a BPS sector containing topologically trivial configurations. This generalizes previous findings, where the BPS sector was populated by kinks (or antikinks), that is, by solutions with $Q=1$ (or $Q=-1$) topological charge. 
This generalization is possible because the solvable self-dual impurity models allow for a spatially non-localized impurity.  In other words, it is not necessary to have $L^2$ integrable impurities. In fact, $\sigma$ can be a constant or even divergent function at $\pm$ infinity. Specifically, we discovered impurities for which the moduli space describes the {\it annihilation} of an initially infinitely separated kink-antikink pair. This kink-antikink pair forms a BPS solution for all values of the moduli space parameter, therefore there is no static force between the constituent solitons. Perhaps the most striking finding is that the full moduli space is a smooth union of solutions  built out  of the regular and singular solutions of the Bogomolny equation of the no-impurity model. They join at the point on the moduli space where the metric takes a completely fine value. 

It is worth noting that the moduli space coordinate, in general,  cannot be associated with the position of the BPS soliton. Especially for the cases with a topologically trivial BPS sector, such an identification is not valid all the time. 

\vspace*{0.2cm}

Looking from a wider perspective, the self-dual impurity models may serve as a very useful tool which allows to decouple the static forces between solitons from the interactions between a soliton and internal modes. For a given process in a given scalar field theory in (1+1) dimensions, we can switch off the static inter-soliton forces by coupling the model with a pertinent impurity in the self-dual manner. Then, we may study the impact of mode-soliton interactions on the dynamical properties of solutions in a very clean set-up. This obviously implies the possibility to go beyond the geodesic approximation, as the interaction with internal modes may strongly affect the dynamics, like, for example, the spectral wall effect. In a next step, the force can be introduced {\it perturbatively} by the addition of an infinitesimally small term which breaks self-duality.   

\vspace*{0.2cm}

There are many directions in which our work can be continued. First of all, one should further study the dynamics beyond the geodesic approximation, which implies the excitation of internal modes on a BPS solution. Recently, this was studied in the topologically non-trivial sector where a mode was excited on an incoming BPS antikink, leading to the discovery of the spectral walls. Now, we can use an impurity such that the moduli space is equivalent to an annihilation process. It is therefore natural to excite a bound mode in this set-up. This can give a chance for a better, analytical understanding of the annihilation of kinks, e.g.,  in $\phi^4$ theory which includes the explanation of the resonant (bouncing) structure as well as  the formation of oscillons. In fact, our framework is model independent and can be applied to any soliton field theory in (1+1) dimensions, see for example \cite{models-1}-\cite{models-6}.

Secondly, any phenomenon discovered in such a setting beyond the geodesic approximation should be relevant also for other self-dual theories in higher dimensions. Here, it is worth to mention the spectral wall effect which should govern the scattering of {\it any} BPS solitons whenever a bound mode enters the continuous spectrum.  In some sense, the spectral wall denotes a {\it forbidden volume} which a sufficiently exited BPS soliton cannot penetrate. If the mode is less exited, the soliton may be temporarily trapped in this volume or, below a critical amplitude of the excitation, easily go through it. 
If a soliton is put in a finite box, internal modes are easily excited, which triggers the spectral wall effect. Of course, the strength of the excitation grows if we increase the number of solitons per volume or, in other words, increase the pressure. Hence, there may exist a phase transition triggered by the formation of spectral walls. Such a phase transition may affect the thermodynamical properties of a BPS multi-soliton system as well as its local, microscopic structure, excluding dynamically a certain volume. Another question is whether such a spectral wall effect survives in near-BPS theories, where static solitons are weakly bound. If this is the case, which is a plausible conjecture, then it may have some implication on near-BPS Skyrme theories \cite{BPS-Sk-1}-\cite{BPS-Sk-3a}, and, as a consequence, on properties of dense nuclear matter.

Finally, a self-duality breaking term should be added as an infinitesimally small perturbation. This leads to the appearance of inter-soliton static forces. Because of the perturbative nature of the deformation, analytical computations should still be possible. This could allow us to disentangle the roles played by mode-soliton interactions and static inter-soliton forces in the dynamics of topological solitons. 

\section*{Acknowledgements}
The authors acknowledge financial support from the Ministry of Education, Culture, and Sports, Spain (Grant No. FPA2017-83814-P), the Xunta de Galicia (Grant No. INCITE09.296.035PR and Conselleria de Educacion), the Spanish Consolider-Ingenio 2010 Programme CPAN (CSD2007-00042), Maria de Maetzu Unit of Excellence MDM-2016-0692, and FEDER.


\begin{thebibliography}{99}

\bibitem{BPS-imp} C. Adam, A. Wereszczynski, Phys. Rev. D98 (2018) 116001. 
\bibitem{BPS-imp-1} C. Adam, T. Romanczukiewicz, A. Wereszczynski, JHEP1903 (2019) 131.

\bibitem{impurity-1} Y. Kivshar, Z. Fei, L. Vazquez, Phys. Rev. Lett. 67 (1991) 1177.
\bibitem{impurity-1a} B. A. Malomed, J. Phys. A25 (1992) 755.
\bibitem{impurity-2} Z. Fei, L. Vazquez, Y.S. Kivshar, Phys. Rev. A46 (1992) 5214. 
\bibitem{imp-barrier-1} Z. Fei, Y. S. Kivshar, L. Vazquez, Phys. Rev. A45 (1992) 6019.
\bibitem{dobrowolski} T. Dobrowolski, Phys. Rev. E65, 036136 (2002).
\bibitem{imp-barrier-2} R. H. Goodman, R. Haberman, Physica D195 (2004) 303.
\bibitem{imp-barrier-3} B. Piette, W. Zakrzewski, J. Phys. A40 (2007) 5995.
\bibitem{imp-barrier-4} J. H. Al-Alawi, W. Zakrzewski, J. Phys. A41 (2008) 315206.  
\bibitem{imp-barrier-5} B. Piette, W. Zakrzewski, Phys. Rev. E79 (2009) 046603.
\bibitem{imp-barrier-6} S. W. Goatham, L. E. Mannering, R. Hann, S. Krusch, Acta Phys. Polon. B42 (2011) 2087. 
\bibitem{step-imp} E. Ekomasov, R. Murtazin, O. Bogomazova, A. Gumerov, J. Magn. Magn. Mater. 339 (2013) 133.
\bibitem{impurity-3} E. Ekomasov, R. Murtazin, V. Nazarov, J. Magn. Magn. Mater. 385 (2015) 217. 

\bibitem{vortex-1} M. Goodman, M. Hindmarsh, Phys. Rev. D52(1995) 4621.
\bibitem{vortex-2} A. A. Izquierdo, W.  G. Fuertes, J. M. Guilarte, Phys. Lett. B753 (2016) 29.
\bibitem{vortex-3} A. A. Izquierdo, W.  G. Fuertes, J. M. Guilarte, JHEP05 (2016) 1.
\bibitem{Sk-vib-1} C. J. Halcrow, Nucl. Phys. B904 (2016) 106.
\bibitem{Sk-vib-2} S. B. Gudnason, C. Halcrow, Phys. Rev. D97 (2018) 125004. 

\bibitem{BPS-imp-2} C. Adam, J. Queiruga, A. Wereszczynski, arXiv:1901.04501.

\bibitem{romanczukiewicz} P. Dorey, A. Halavanau, J. Mercer, T. Romanczukiewicz, Y. Shnir, JHEP 1705 (2017) 107.

\bibitem{BPS-imp-DW} C. Adam, K. Oles, T. Romanczukiewicz, A. Wereszczynski, arXiv:1902.07227.

\bibitem{SM} N. Manton, P. Sutcliffe, {\it Topological Solitons}, Cambridge University Press, Cambridge, 2004.
 
 \bibitem{dunne} G. Dunne, J. Feinberg, Phys. Rev. D57 (1998) 1271.
\bibitem{BPS-imp-SW} C. Adam, K. Oles, T. Romanczukiewicz, A. Wereszczynski, arXiv:1903.12100.

\bibitem{leese} R. A. Leese, Nucl. Phys. B334 (1990) 33. 

\bibitem{cp1-zakrzewski} B. Piette, W. Zakrzewski, Nonlinearity 9 (1996) 897.

\bibitem{Zakrzewski} W. Zakrzewski, {\it Low Dimensional Sigma Models}, Bristol, Institute of Physics Publishing, 1989.
 
 \bibitem{schroers-1} B. Barton-Singer, C. Ross, B. Schroers, arXiv:1812.07268.
 
  \bibitem{schroers-2} B. Schroers, arXiv:1905.06285.
  
\bibitem{han} X. Han, G. Huang, Y. Yang, arXiv:1902.09448.

\bibitem{weigel-1}  I. Takyi, H. Weigel, Phys. Rev. D94 (2016)  085008. 

\bibitem{weigel-2} H. Weigel, arXiv:1809.03772.

 \bibitem{models-1} V.A. Gani, M.A. Lizunova and R.V. Radomskiy, JHEP04 (2016) 043.
 
 \bibitem{models-2}  A. Alonso-Izquierdo, Phys. Rev. D97 (2018) 045016. 
 
 \bibitem{models-2a} D. Bazeia, R. Menezes, D.C. Moreira, J. Phys. Comm. 2 (2018) 055019. 
 
 \bibitem{models-3} L.A. Ferreira, P. Klimas and W.J. Zakrzewski, JHEP01 (2019) 020.
 
 \bibitem{models-3a} D. Bazeia, A. R. Gomes, K.Z. Nobrega, F. C. Simas, Phys. Lett. B793 (2019) 26. 
 
 \bibitem{models-4} I C. Christov, R. J. Decker, A. Demirkaya, V. A. Gani, P.G. Kevrekidis, A. Khare,  Phys. Rev. Lett. 122 (2019) 171601.  
 
 \bibitem{models-5} N.S. Manton, J.Phys. A52 (2019) 065401.
 
 \bibitem{models-6} I. C. Christov, R. J. Decker, A. Demirkaya, V. A. Gani, P.G. Kevrekidis, R.V. Radomskiy, Phys. Rev. D99 (2019) 016010.

\bibitem{BPS-Sk-1} C. Adam, J. Sanchez-Guillen, A. Wereszczynski, Phys. Lett. B691, 105 (2010).
\bibitem{BPS-Sk-1a} C. Adam, C. Naya, J. Sanchez-Guillen, A. Wereszczynski, Phys. Rev. Lett. 111 (2013) 232501.
\bibitem{BPS-Sk-2} P. Sutcliffe, JHEP 1008 (2010) 019.
\bibitem{BPS-Sk-2a} C. Naya, P. Sutcliffe, Phys. Rev. Lett. 121 (2018) 232002.
\bibitem{BPS-Sk-3} M. Gillard, D. Harland, M. Speight, Nucl. Phys. B 895 (2015) 272. 
\bibitem{BPS-Sk-3a} S. B. Gudnason, Phys. Rev. D98 (2018) 096018.


 



\end{thebibliography}
\end{document}